# Damage of Cross-Linked Rubbers as the Scission of Polymer Chains: Modeling and Tensile Experiments

Alexei Y. Melnikov and A.I. Leonov

Department of Polymer Engineering, The University of Akron
Akron, Ohio 44325-0301, U.S.A.

**Abstract**

This paper develops a damage model for unfilled cross-linked rubbers based on the concept of scission of polymer chains. The model is built up on the well-known Gent elastic potential complemented by a kinetic equation describing effects of polymer chain scission. The macroscopic parameters in the damage model are evaluated through the parameters for undamaged elastomer. Qualitative analysis of changing molecular parameters of rubbers under scission of polymer chains resulted in easy scaling modeling the dependences of these parameters on the damage factor. It makes possible to predict the rubber failure in molecular terms as mechanical de-vulcanization. The model was tested in tensile quasi-static experiments with both the monotonous loading and repeated loading-unloading.

**Introduction**

The damage in materials is commonly defined as micro-failure caused by formation of micro-voids and/or micro-cracks in intense mechanical fields. In polymers, formation of microvoids is always caused by scission of covalent bonds between monomers in macromolecules.

During almost sixty years of studies, failure properties of rubbers were mostly investigated in simple extension experiments. Pioneering contribution was made by P. Flory and co-workers [1] who studied failure of natural rubbers in slow (quasi-static) tensile experiments. Later on, T. Smith [2, 3] discovered the "failure envelope". He found that the stress-strain behavior at failure also depends on the rate of extension and presented the data as time-temperature superposed plots.

Several molecular theories of ultimate behavior of rubbers at failure have also been developed almost at that time and were favorably compared with experiments [4, 5]. Since typical dissociation energy of (mostly, sulfur) cross-links is only slightly lower than that for covalent bonds in polymeric chains [4] the difference between these two energies were ignored in the theories [4, 5]. Their results were widely exposed later in the texts [6-8]. These texts also discussed many notched, crack initiated experiments, including tearing of rubber sheets, which used for evaluations of engineering properties of elastomers.

Early molecular models could not, however, analyze accumulating the material microscopic defects in polymers under various loading conditions. These problems have been analyzed using various damage approaches. Many versions of the damage approach for rigid materials with small elastic strains have been developed and tested. We refer here only to the recently developed mathematical models [9-11]. The damage in cross-linked rubbers and gels caused by scission of polymer chains is usually irreversible. In mechanical interpretation, it means that after unloading, the previously damaged material does not regain its initial mechanical properties. It should be mentioned that completely elastic continuum damage approaches [12, 13] for filled rubbers have also been developed to describe the Mullins effect in elastomers using irreversible thermodynamics. These approaches operated with several fitting parameters (4 in paper [13]). However, in many cases the Mullins hysteresis in rubbers is completely reversible (see discussions in Refs. [13], [14]), which questions the application of the irreversible approach.

The common formulation of damage theories in continuum mechanics is as follows: (i) choosing a constitutive equation capable to describe the mechanical properties of undamaged material, (ii) introducing the "damage factor" as the micro-crack/void volume concentration $d$, with the free energy and stress of undamaged material multiplied by the damage factor $\chi = 1 - d$, (iii) formulation of equation describing evolution of damage factor, and (iv) formulation of a local failure criterion. The key steps (ii) and (iii) in the damage theory have been seemingly first introduced in paper [11].

Many papers today analyze together the damage and fatigue failure in various materials. The calculations and tensile tests of a damage approach for fatigue failure of filled natural rubber are illustrated in Ref. [15]. Here, as in paper [12], the constitutive modeling of undamaged rubber was described by the Ogden approach [16], the damage kinetics using the approach [11], and the failure criterion was taken as $d = 1$.

The results of damage simulations of low amplitude fatigue, obtained in the last century, have been thoroughly compared with experimental data in the review paper [17]. The attention was paid to distinguish the difference between the crack nucleation and propagation. Other papers also employed numerical studies of 2D or 3D problems of rubber failure using various continuum damage approaches. Some of them also compared the results of calculations with data.

The beginning of the 21$^{st}$ century has been marked with a burst in publications on damage and fatigue of rubbers. Plenty of papers with new theoretical continuum approaches have been published in *Mechanics and Physics of Solids*, *International Journal of Fatigue*, and *Mechanics of Materials*. Among general results, few additional effects have also been discussed. The most important is the effect of microscopic voids formed during the damage of rubbers. If the voids are so small that they are stabilized by surface tension, the applied macroscopic stretches could break the void stability causing their cavitations and material failure. These voids may serve as precursors of cracks, which occur as the final result of evolution of the voids. This effect was first described and analyzed long ago [18] (see also more recent paper [19]). Recent paper [20] analyzed the voids formation and their evolution in tensile and torque tests of rubbers.

Incorporating the effect of voids into the damage approach can result in non-traditional damage theory [21] with compressibility effects for the whole deformable body. This is different from the standard incompressible damage models. The incompressibility assumption in the majority of damage models might be justified if the concentration of damaged structure elements (e.g. broken macromolecules) is considerably higher than the volume concentration of voids, which in turn is insignificant as compared with the free volume of polymers, almost up to macroscopic failure of a sample.

Another is effect of very fast (supersonic) loading on rubber rupture analyzed in paper [22]. The chemical effect of rubber oxidative ageing was analyzed using a damage approach in paper [23]. Effects of thermo-oxidation on strengths of rubber vulcanizates have been discovered long ago [24]. Finally, the effect of thermal fatigue caused by viscoelastic effects, well understood for plastics (e.g. see Ref. [25]), seems to be inapplicable to low amplitude fatigue of such soft materials as unfilled cross-linked rubbers, at least when the frequencies are not extremely high.

Other ideas for describing scission of chains in polymers have been developed in series papers (see references in paper [26]), based on continuum mechanistic approach with fitting functions. These models have never been tested experimentally.

A choice of constitutive equation (CE) for undamaged material can be based on the fact that the entropic effects overwhelmingly contribute in equilibrium physics and mechanics of rubber deformations. For the low and modest strains they have been well described long ago [27] using simple Gaussian statistics. Direct non-Gaussian statistical calculations up to the ultimate extension of polymeric chains in cross-linked rubbers [28] resulted in awkward formulations and were not compared to the data. Therefore simplified descriptions of entropic behavior of rubbers under ultimate extension have also been developed [29-31]. Both the entropic and enthalpic contributions in nonlinear rubber elasticity were recently analyzed in paper [32] where the effect of strain-induced crystallization in rubbers was also modeled. It was shown that including enthalpic contributions in stress removes singularity of free energy and stress at very high extensions. Evidently, the enthalpic effects, related to the microscopic distortions of rigid covalent bonds, are extremely important for molecular evaluations of scission of polymer chains. Using numerical calculations based on the density functional theory, paper [33] reported the values of forces, energies and extension of bonds, contributing in enthalpic type of rubber elasticity with scission of covalent C-C bonds of isoprene and butadiene molecules. Employing these results, some evaluations of tensile strength of respective rubbers have also been discussed [33].

Not so much attention has been paid to molecular modeling of viscoelastic effects in cross-linked unfilled rubbers. In non-steady deformations, these effects are caused by the molecular friction of polymer chains between cross-links. A discovery of continuous change in rubber relaxation properties in course of vulcanization [34] made possible to use even in the solid, cross-linked state the "liquid like" nonlinear modeling of relaxations. A good example of the latter is thermodynamically related, differential type of CE's well established for polymer melts and green rubbers [35, 36]. The additive modeling of equilibrium and non-equilibrium stresses in cross-linked rubbers have been proposed and discussed in several papers. Ref. [37] exemplifies using the nonlinear differential type of relaxation modeling. The additive approach with nonlinear integral type modeling of relaxations was started with BKZ models [38, 39] and was recently re-visited along with some new experimental studies [40].

All the cited above models of solid rubber viscoelasticity as well as many others proposed in literature, are phenomenological. Yet there was a recent attempt [41] to develop a molecular type theory of viscoelasticity for cross-linked rubbers. It is based on Rouse-like models for polymer chains with constraints caused by cross-links. In spite of promising character, the relaxation function in linear limit has not been calculated or modeled in paper [41].

In general the mechanical energy supply in elastomers, which leads to both deformations and damage, consists of elastic and viscoelastic contributions. Modeling of these two parts based on special experimental studies is described in Refs. [42,43] where viscoelastic part was modeled due to the papers [35-37]. It was found that the relaxation moduli $G_k$ in cross-linked rubbers are at least an order of magnitude lower than the equilibrium modulus $G_e$ [42, 43]. Therefore in slow enough deformations the relaxation effects due to molecular friction could be neglected.

**Continuum Damage Model**

The phenomenological model developed below is based on the idea that the energy for chain scission is supplied from the mechanical energy $W$ accumulated in the elastomer deformations. In this paper we consider slow, quasi-static deformations as pseudo-elastic and neglect contribution of molecular relaxations.

Introducing the damage parameter is based on the two physical assumptions:

(i) In cross-linked elastomers, the scission of polymer chains caused by dissociation of covalent bonds is irreversible. This is because the free radicals formed in the act of scission, are quickly neutralized [33], typically by oxidation.

(ii) The volume concentration of microscopic voids formed by the scission of polymer chains is negligible almost until the failure.

The assumption (i) means that in any unloading the damage accumulated during previous loading of material, is conserved and memorized. It can only increase with the increase of loading. The assumption (i) also means that the regularity with which the loading-unloading procedure affects the rubber is not essential. The assumption (ii) makes possible to apply the incompressibility condition which is commonly in use for damage models.

It is convenient to introduce the damage factor as the volume concentration of unbroken bonds $\chi = 1 - d$ $(0 < \chi < 1)$. The elastic damage model developed below correlates the damage parameter $\chi$ to the strain/stress variables.

The main problem with modeling broken polymer chains in soft cross-linked rubbers is that the covalent bonds can be broken only after a significant extension of the chains. It means that the free energy of rubber is mostly spent initially on stretching and orientation of stretched macromolecules. Since statistics of chain orientation is not developed in this paper, we present below an approximate approach [42-44] for rough modeling of damage process. In this modeling, the visible breakage of polymer chains is assumed to begin after accumulation of an average macroscopic energy $U_s$ to extend and orient the chains and prepare them for scission. Here the value of $U_s$ is considered as a phenomenological parameter. Nevertheless, it is possible to evaluate $U_s$ through molecular parameters using the results of our paper [32], bearing in mind that the breakage of covalent bonds starts when the enthalpic contribution in rubber elasticity is getting significant [33].

If the level of mechanical energy $W$ supplied to the system from unbroken chains is less than the threshold value $U_s$, the rubber is considered as undamaged ($\chi = 1$). After that, the damage increases ($\chi$ decreases) due to the difference $\chi W - U_s$. Using this approximation, the general kinetic equation for the damage factor $\chi$ was proposed in Refs. [42-44] as energy balance equation:

$$-U_s \frac{d\chi}{dt} = \chi \frac{d}{dt}(\chi W - U_s); \quad \chi|_{t=0} = \chi_0. \quad (1)$$

The left-hand side of (1) presents the rate of chain scission while the right-hand side the rate of energy supply from unbroken chains. The known value $\chi_0$ ($\leq 1$) is the initial damage accumulated in previous loadings. The particular case $\chi_0 = 1$ describes the deformation of non-damaged, fresh material.

Integrating (1) yields:

$$\chi = \begin{cases} \chi_0 e^{-(P-1)}, & P > 1 \\ \chi_0, & P \leq 1 \end{cases} (\dot{P} > 0), \quad \chi = const \ (\dot{P} \leq 0); \quad P = \frac{\chi W}{U_s}. \quad (2)$$

The inequalities imposed in (2) guarantee the monotonous decrease of damage parameter $\chi$ for any, generally non-monotone loadings. In case of quasi-static deformations, considered in this paper, $W(T, \chi, I_1, I_2)$ is elastic pseudo-potential, $I_1 = tr\underline{\underline{B}}$ and $I_2 = tr\underline{\underline{B}}^{-1}$ are two basic invariants under a incompressibility assumption, $I_3 = \det \underline{\underline{B}} = 1$, and $\underline{\underline{B}}$ is the Finger tensor. Equation (2) is coupled with the stress-strain equations for nonlinear elastic solids affected by damage,

$$\underline{\underline{\sigma}}(\chi, \underline{\underline{B}}) = -p\underline{\underline{\delta}} + 2\underline{\underline{B}} \cdot \partial W / \partial \underline{\underline{B}}\big|_{T, \chi}. \quad (3)$$

Equations (2)-(3) are the closed set, as soon as expression for $W(T, \chi, I_1, I_2)$ is specified.

Involving the damage factor $\chi$ in expression for elastic pseudo-potential $W$ is a crucial step in developing damage models. In typical damage theories the dependence $W$ on $\chi$ is formulated *ad hoc* within the common formalism of irreversible thermodynamics. It involves in formulations several fitting parameters. Another way of modeling employed in this paper is using some qualitative ideas of changing molecular parameters during polymer damage, caused by scission of polymer chains. Using this approach allows to avoid involving fitting parameters in mathematical formulation.

The proposed model (2)-(3) is dissipative, with equation (1) describing the local rate of dissipation, i.e. the lost energy due to the irreversible scission of chains. The measurement of dissipation heat in rubber damage process was reported in paper [44].

To specify the elastic potential $W_e(T,\underline{\underline{B}}) = W(T,\underline{\underline{B}},\chi)|_{\chi=1}$ for undamaged elastomer we should make a choice from the most popular elastic CE's [29-31]. All of them are of entropic type. It was found [46] that the different CE's [29-31] describe the rubber mechanics almost with the same precision. Therefore the preference was given to the simplest Gent potential [30]:

$$W_e(T,\underline{\underline{B}}) = G_e(T)\frac{I_*-3}{2}\ln\left(\frac{I_*-3}{I_*-I_1(\underline{\underline{B}})}\right); \quad \underline{\underline{\sigma}} = -p\underline{\underline{\delta}} + G_e\frac{I_*-3}{I_*-I_1}\underline{\underline{B}}; \quad I_1(\underline{\underline{B}}) = tr\underline{\underline{B}}. \quad (4a)$$

Parameter $I_*$ describing ultimate stretching of strands was treated in [30] as fitting one, being in order of 100. The simplicity of potential (4a) was seemingly the reason why it was also used in the damage model [13]. It should be noted, however, that the potential (4a) could be used with reservations, because in general the behavior of elastomers distinctly depends on the second invariant $I_2$ [27].

For simple extension formulas (4a) take the form:

$$W_e = G_e\frac{I_*-3}{2}\ln\left(\frac{I_*-3}{I_*-I(\lambda)}\right), \quad I(\lambda) = \lambda^2 + 2/\lambda, \quad \sigma = G_e\frac{I_*-3}{I_*-I(\lambda)}(\lambda^2 - 1/\lambda). \quad (4b)$$

**Evaluations of Molecular Parameters**

*3.1. Molecular Evaluations of Parameters in Potential for Undamaged Elastomer*

The elastic model (4) has two parameters - the shear modulus $G_e$ and parameter $I_* \approx \lambda_*^2$, where $\lambda_*$ corresponds to the ultimate stretching of polymeric chain. In case of moderate strains when $I_1 \ll I_*$ equation (4a, b) coincides with classical theory of rubber elasticity, so the elastic modulus $G_e$ is presented as $G_e = \nu_c kT$. Here $k$ is the Boltzmann constant and $\nu_c$ is the number of cross-links per unit volume, given by the Flory's formula [47]:

$$\nu_c = \frac{\rho A}{M_c}\left(1 - 2\frac{M_c}{M_n}\right) \quad (5)$$

In (5) $\rho$ is density, $M_c$ is the average molar mass of polymer chains ("strands") between crosslinks in non-damaged rubber, $A$ is the Avogadro number and $M_n$ is the number averaged molecular mass of polymer chain before vulcanization. The bracket in the right-hand side of (5) accounts for effect of chain free ends. It is usually ignored in rubber theories because its value is close to unity in common case of soft elastomers when $M_c/M_n \ll 1$. The rubber elastic modulus is then presented as:

$$G_e = \frac{\rho RT}{M_c}\left(1 - 2\frac{M_c}{M_n}\right), \quad (6)$$

where $R$ is the gas constant. The value $M_c$ is commonly evaluated from (6) if $M_n$ and $G_e$ are known from measurements.

We now present a possible way of expressing $I_*$ via molecular parameters. It starts with estimation of the ultimate stretching of a single strand using the single-molecular approach common in the theory of rubber elasticity [27] and then takes into account the effect of cross-links connectivity. To evaluate the stretching simple strand we use a common modeling of molecular structure of unloaded cross-linked rubbers as a set of coil-like molecular strands (Fig.1a). Here each polymer coil contains average number of $P_c = M_c/m$ monomers of size $l_m$, where $m$ is the molecular mass of monomer unit. The size of coil $d_c$ is estimated as the diameter of gyration of ideal chain, amended for real chain as $d_c = 2\sqrt{<g>} = 2Kl_m\sqrt{P_c/6}$, where the Kuhn factor $K$ is calculated using statistical theory of macromolecular chains [48]. The totally extended length of coil is: $l_{*c} = P_c l_m$. The macroscopic ultimate stretching ratio is then approximated as $\lambda_{*c} = l_{*c}/d_c = \sqrt{6P_c}/(2K)$. Thus the value of $I_{*c}$ for a single chain is calculated in the form:

$$I_{*c} \approx \lambda_{*c}^2 = \frac{3P_c}{2K^2}. \tag{7}$$

Here the values of molecular parameters in right hand side of (7) are known [48]. Formula (7) describes, however, the ultimate stretching of a single strand and does not take into account the connectivity of the strands caused by the cross-links. This effect is particularly important at ultimate stretches, when two strands merge in one with average molecular mass equal to $M_c^* \approx 2M_c$. More generally $M_c^* \approx fM_c/2$, where $f$ ($\approx 4$) is the average functionality of cross-links. Then in the continuum approach one can use formula (7) with the change $P_c \to fP_c/2$, which gives the final formula for evaluation of $I_*$ as:

$$I_* \approx \lambda_*^2 \approx (f/2)\lambda_{c*}^2 = \frac{3fP_c}{4K^2} = \frac{3fM_c}{4K^2m} \tag{8}$$

This formula presents the ultimate deformation in molecular network only through the molecular parameters of rubber chains. Alternative "blob" approach to evaluate $I_*$ is explained in Fig.1b. Here an average blob, representing a polymer coil centered at cross-link, consists of $f/2$ cross-linked strands. Assuming that the gyration radius of blob is approximately the same as for single coil, one can obtain (7) once again using the condition of equivalency for two entropy descriptions of blobs and strands. Here the number of strands is $f/2$ times higher than the number of blobs.

*3.2. Change in Molecular Parameters of the Elastic Model with Damage*

We now make assumptions of changing rubber molecular parameters with scission of polymer chains. Evidently, increasing damage (decreasing $\chi$) will increase $M_c$, decrease $f$, and decrease $v_c$. Note that increasing $M_c$ because of the scission of polymer chains has been proposed earlier in paper [49] and also used in Ref. [50] to describe the Mullins effect. Consider now the term $2M_c/M_n$ which characterizes the contribution of concentration of chain ends in the cross-link density $v_c$. There are two synergetic contributions in change of this term caused by scission of polymer chains. The first one is direct increase in the term $2M_c/M_n$ with decreasing $\chi$ which was simply modeled as $\chi^{-1}$ because the number of end chains increases in any act of chain scission. The second one is additional increasing in value of $M_c$ with decreasing $\chi$. Assuming these dependences in a simplest, scaling way yields the following rude modeling:

$$\hat{M}_c(\chi) \approx M_c/\chi, \quad \hat{f}(\chi) \approx f\chi, \quad \hat{v}_c(\chi) = \frac{\rho A \chi}{M_c}\left(1 - \frac{2M_c}{\chi^2 M_n}\right). \tag{9}$$

Here the molecular characteristics of elastomer changed with scission denoted by upper tilde. The physical reason for the first two formulas in (9) is the releasing macromolecules from cross-links by either seldom scission of cross-links or by overwhelming scissions of surrounding them chains.

The last expression in (9) is obtained using the first formula in (9) and assuming that the Flory formula (5) is valid for damaged elastomers, as the damaged molecular network were assembled from the damaged macromolecules. When $v_c(\chi) \to 0$, this relation predicts the failure due to the scission of polymer chains as an effective de-vulcanization or mechanical degradation of polymer network. It means that at this limit the cross-linked rubber solid is converted in non-cross-linked (green) rubber, which unlike the fresh material is highly branched. The critical concentration of unbroken bonds in this limit is:

$$\chi^* = \sqrt{2M_c/M_n} \tag{10}$$

This limit can be treated as the molecular criterion of macroscopic failure due to de-vulcanization. Evidently, this criterion cannot completely describe the failure of rubber sample, because other issues as micro-void accumulation converted to micro-cracks, and magisterial crack propagation affect the rubber behavior close to the failure. Yet we hope that this rough criterion will be close to the observable failure of rubber.

Using (6) along with (9), we now can evaluate the change in two basic mechanical parameters $G_e(T, \chi)$ and $I_*(\chi)$ in the damage process as:

$$\hat{G}_e(\chi) = \frac{\rho RT \chi}{M_c}\left(1 - \frac{2M_c}{\chi^2 M_n}\right), \quad \hat{I}_*(\chi) \approx \frac{3\hat{f}(\chi)\hat{M}_c(\chi)}{4K^2 m} = \frac{3fM_c}{4K^2 m} = I_* \tag{11}$$

Formulas (11) establish the dependence of elastic pseudo-potential $W$ on the damage factor $\chi$, when substituting in (4) $G_e(\chi)$ from (11) instead of $G_e$. The second relation in

(11) demonstrates the independence of ultimate strain parameter $I_*$ from the damage parameter $\chi$, resulted from our rough scission modeling (9).

Additional effect of strain induced crystallization for crystallizable rubbers should also be involved in the model, especially in simulations of tensile experiments for natural rubber (NR). A simplified modeling of this effect was proposed in [31] as:

$$\hat{G}_e(\chi) \to \hat{G}_e(\chi)/(1-\kappa_c), \quad \kappa_c(\lambda) \approx \alpha_c \frac{\lambda - \lambda_0}{\lambda_f - \lambda_0} \quad (\lambda_0 \leq \lambda \leq \lambda_f). \tag{12}$$

Here $\alpha_c$ is the ultimate (maximal) degree of stress-induced crystallinity, depending on molecular mobility and cross-link concentration $v_c$; $\lambda_0$ and $\lambda_f$ are respectively the stretching ratios at the beginning and end of crystallization. According to paper [51] the strain-induced crystallization did not show a significant accumulation of crystals in repeating loadings. Therefore one can assume that in repeating elongations the "freshly made" crystals completely disappear after unloading.

Thus the final set of equations for elastic damage theory consists of kinetic equation (2) and the elastic relations (4) modified for the damage and possible strain induced crystallization. These modifications are shown below for simple extension as:

$$W = \hat{G}_e(\chi)\frac{I_* - 3}{2}\ln\left(\frac{I_* - 3}{I_* - I(\lambda)}\right), \quad I(\lambda) = \lambda^2 + 2/\lambda, \quad \sigma = \hat{G}_e(\chi)\frac{I_* - 3}{I_* - I(\lambda)}(\lambda^2 - 1/\lambda). \tag{4.c}$$

Here $\hat{G}_e(\chi)$ is given in (11), and in case of crystallizable rubbers one should also use (12).

**Tensile Experiments and Comparisons with Model Calculations**

*Experimental*

The base material for our tensile experiments was natural rubber (NR), *cis*-1,4-polyisoprene with almost standard curing additives for sulfur accelerated vulcanization. The samples from non-vulcanized (green) NR plates were shaped in the RAM press with following vulcanization in optimum at $140\,^0\mathrm{C}$ during 30 minutes.

The standard Instron 5567 tensile tester with Series IX version 8.13.00 software was employed in our experiments. It basically worked in two modes – extension with a constant linear speed $u$, and relaxation with following rest. The lower speed of extension was $u \approx 1$ mm/min, and highest possible linear speed $u \approx 900$ mm/min. All the tests were undertaken at the room temperature.

We found that the common dog-bone samples for the tested NR demonstrated very inhomogeneous strain field. Additionally, the material began chipping out of clamps at moderate strains. Our attempts to increase the clamp pressure resulted in tearing the dog-bone samples in clamps. To avoid these detrimental effects we used in our experiments O-ring samples with hooks ("holders") (Fig. 2). The initial diameter of round cross-section of each O-ring was equal to 1 cm.

Since initial deformation in these samples was mostly related to straitening of two branches of O-ring samples, the strains in the branches were established by measuring cross-sections in the section of homogeneous extension. After small enough stretching ratio $\lambda \approx 1.06$, where the true strain and stress were estimated by a change in cross-sections, the strain fields in two branches of O-ring samples were highly homogeneous, except relatively small regions near the holders. Therefore the stretching ratio for monotonous extension was calculated as $\lambda(t) = k + ut/L_0$, where $k$ and $L_0$ were found empirically. With the value $L_0 \approx 11.7 cm$, the maximum lower $\dot{\varepsilon}_{low}$ and higher $\dot{\varepsilon}_{high}$ extension rates were equal to $\dot{\varepsilon}_{low} \approx 6.8 \times 10^{-4} \sec^{-1}$ and $\dot{\varepsilon}_{high} \approx 0.128 \sec^{-1}$, respectfully. In the both ultimate cases the $\sigma - \lambda$ extensional plots almost coincided.

The true elongation stress $\sigma(t)$ in samples was determined through the measured elongation force $F(t)$, which made possible to establish for the small $u$ values the quasi-static dependence $\sigma(\lambda)$. These experiments allowed us to achieve the ultimate values of $\lambda \sim 7$ and more, up to the breakage of sample. It should be noted that the breakage always happened near a holder, where the stress and strain fields were inhomogeneous.

Experiments and calculations for larger speeds of extension showed their dependence on the loading speed $u$, thus revealing viscoelastic contributions in stress. In this case the relaxation effects, however small, are visible. These experiments and their viscoelastic modeling have been demonstrated in almost not damaging region of tensile strains in our paper [42], and thesis [43]. These references also showed the results of measuring the linear viscoelastic spectrum obtained in separate shearing experiments. In particular, the value of equilibrium shear modulus $G_e \approx 0.36$ MPa found in these shearing experiments was also confirmed in our tensile experiments.

In order to find the region of deformations where the damage effects could be neglected, we undertook a large series of loading - unloading experiments. Some of these experiments are shown in Figs.3a, b for repeating slow quasi-static loading. They show that in the region $\lambda < 3.5$ ($\approx 250\%$) the repeated loading curves almost coincide with precision $\pm 7\%$. The same results were confirmed for the fast loading with following relaxation.

*Numerical Values of Parameters*

Two parameters describe the mechanics of undamaged elastomers presented by Gent CE (4) - the shear modulus $G_e$, and parameter $I_*$ characterizing the limit strain in entropic approximation of rubber behavior. These parameters are presented above via molecular characteristics of elastomer $M_n, M_c, f, m$, and $K$ by the well-known formula (6) and expression (8) derived in this paper. These formulas shown for convenience together are:

$$G_e = \frac{\rho RT}{M_c}\left(1 - 2\frac{M_c}{M_n}\right), \quad I_* \approx \frac{3 f M_c}{4 K^2 m} \quad (13)$$

Here $\rho$ is density, $R$ is the gas constant, $T$ is the Kelvin temperature, $M_n$ is the number average molecular mass of pre-vulcanized rubber macromolecules, $M_c$ is the average molecular mass of part of macromolecules between cross-links, $f$ is the average coordinate number of cross-links, $m$ is the molecular mass of monomer units in polymer chains, and $l_m K$ is the length of the Kuhn's segment (correlation length) along the polymer chain. Unlike other parameters in (13) known from literature, parameter $M_c$ is commonly found from independent macro-experiments. The easiest way is to establish its value from measurement of modulus $G_e$ at small strains.

We use in the following the literature data for molecular parameters of natural rubber (NR), *cis*-1,4-polyisoprene. The value $M_n = 138,000$ after mastication for our samples, measured by GPC using tetrahydrofuran as a solvent, is close to the value $M_n \approx 100,000$, estimated for NR in paper [1]. The length of monomer link

$l_m = 5.05 \text{Å}$ and molecular monomer mass $m = 68$ were found from the data for polyisoprene in Ref. [48] (Chapter 5). Using the experimental data for our samples at small strains, we found the value of $G_e = 0.364$ MPa. Then due to the first formula in (13) the molecular mass $M_c$ of chain between the cross-links is: $M_c \approx 6200$. It means that the average degree of polymerization of chains between cross-links is $P_c = M_c / m \approx 91$. From the Table 11 in Chapter 5 of the text [48] we can also find that $K^2 = 4.7$, i.e. the value of the Kuhn parameter $K \approx 2.17$. Additionally, the elastomer network is commonly assumed as tetrafunctional, i.e. $f = 4$. Then the value of parameter $I_* \approx 58$ was calculated using the second formula in (13) derived in this paper.

Unlike great majority of synthetic elastomers, the tested samples from NR display in simple extension the strain induced crystallization. This effect, roughly described in paper [39], was also taken into account in tests of our model. The values of empirical parameters $\lambda_0 = 4$, $\lambda_f = 7$, $\alpha_c = 0.28$ in the coefficient $k_c(\lambda)$ in (12) were found by fitting experimental data in Ref. [27], p.21.

Finally, the key value $U_s$ for the damage kinetics in formulas (1) and (2) is approximately evaluated as an effective damage threshold in simple extension, corresponding to the stretching threshold $\lambda_c$ before which the damage is neglected, i.e. $\chi = 1$ if $\lambda < \lambda_c$. As it was shown in Figures 3a, b - $\lambda_c \approx 3.5$. At this value of $\lambda$, the strain induced crystallization effect is negligible. According to (2) in this approximation,

$$U_s \approx W_e(\lambda_c).  \qquad (15)$$

Using formula (4b) for $\lambda = \lambda_c = 3.5$ and previously found values of parameters $G_e = 0.364$ MPa and $I_* \approx 58$, the calculation yields $U_s \approx 1.91 MPa$. Remarkably, the value $\lambda_c \approx 3.5$ was also recently reported as the beginning of distortion of covalent bonds (enthalpic effects) in paper [33], which described computational physics simulations of isoprene scission.

The values of parameters determined in this Subsection are shown in Table 1.

Table 1: Values of parameters

| $G_e$=0.364 MPa (13) | | $I_* = 58$ (13) | | | $k_c(\lambda)$ (12) | | | $U_s = W_e(\lambda_c)$ (15) | |
|---|---|---|---|---|---|---|---|---|---|
| $M_n$ | $M_c$ | $m$ | $f$ | $K$ | $\lambda_0$ | $\lambda_f$ | $\alpha_c$ | $\lambda_c$ | $U_s$ |
| 138,000 | 6,200 | 68 | 4 | 2.17 | 4 | 7 | 0.28 | 3.5 | 1.91MPa |

In this table, the macroscopic parameters of the model $G_e$, $I_*$, $U_s$ with formula numbers they were used, and those in the function $k_c(\lambda)$ in (12), are shown in the upper row of Table 1, just above the corresponding values of detail parameters demonstrated in the Table.

Note that all the values of these detailed parameters have been obtained either from literature sources or from independent experiments. It means that no fitting parameter is involved in comparison of our model with tensile experiments in the damage region of extension.

It is also of interest to compare the value $U_s$ in Table 1 with the tabulated values [52] of dissociation energies for C-C bonds ($E_{C-C}$ = 347kJ/mol), C-S bonds ($E_{C-S} = 272$ kJ/mol), S-S bonds ($E_{S-S} = 265$ kJ/mol), and C=C bonds ($E_{C=C} = 614$ kJ/mol) (see more detailed data in Refs. [53, 54]). As in paper [4], we neglect the difference between the energies of S-S (C-S) and C-C bonds. We also neglect as improbable the dissociation of high energy C=C bonds. Then an easy calculation shows that for our NR samples, the average scission energy $U_s$ is approximately 20 times less than the value of $E_{C-C}$. It means that on average only 1/20 (or ~ 4.5) monomer bonds between the cross-links can be ruptured.

*Comparison of Model Calculations with Tensile Experiments*

Figures 4-8 demonstrate comparison of our tensile experiments for NR with modeling calculations based on equations (2), (4c) and (12). The values of parameters are shown in Table 1. The computational aspects were easy and we used MATLAB for calculations.

Figure 4 compares the calculated dependence $\sigma(\lambda)$ for fresh material up to the sample break with experimental data (red line), while Figure 5 demonstrates the calculated plot of damage factor $\chi$ versus stretching ratio $\lambda$. Two regions of visible deviations between calculated and experimental curves are seen in Figure 4. The first one around $\lambda = \lambda_c = 3.5$ is due to our rude modeling of damage beginning (Fig.5). The second one is in the region of high values of $\lambda$ near the rupture, where the Gent CE (as well as any other CE discussed before) is singular when ignoring the enthalpic contribution in stress. Because of this reason the calculated value of stress at rupture $\sigma^* \approx 238\,\text{MPa}$ is much higher than the experimental one $\sigma^*_{ex} \approx 46\,\text{MPa}$.

Figures 6 and 7 demonstrate the behavior of preliminary damaged NR. The Figure 6 presents the first, etalon extension curve $\sigma(\lambda)$ for a fresh NR sample up to $\lambda_1 = 5.8$, while the Figure 7 the second extension up to $\lambda_2 = 5.7$ of damaged sample with loading history shown in Figure 6. The unrecoverable deformation was taken into account in the stretching ratio calculations for the second cycle. In both figures the curves shown by symbols and red colored lines denote the calculations and experiments, respectively. The primary and secondary extensions were shown in different figures because of scattering the data. It is seen that the stress-strain curves in Figure 7 are essentially lower than those in the Figure 6. The third and other consecutive cycles of repeated deformations for the NR sample, damaged in the second cycle, are almost indistinguishable from the second one. The reason for this is the stabilization of damage process in higher numbers of cycles, demonstrated in Figure 8.

Generally, Figures 4-8 demonstrate a good enough agreement of our model predictions with experimental data. Especially important here is a successful description of the second loading of damaged sample.

The ultimate value of damage factor calculated due to (10) is $\chi^* \approx 0.3$. It is almost 25% lower than the ultimate value of $\chi_{cr} \approx 0.4$ at rupture of sample. It might seem that the real breakage of sample happens earlier than the complete mechanical de-vulcanization of elastomers predicted by formula (10), seemingly due to the cracking of the sample. It could also be explained by mentioned above macroscopic rupture samples near a holder where there stress concentration involves higher damage (lower value of $\chi_{cr}$) as compared to its value calculated for homogeneous extension. The last explanation might improve our prediction of damage factor at sample breakage.

It should also be finally mentioned that the calculations of the $\sigma - \lambda$ plots using the damage model with and without the multiplier $1 - 2M_c/(\chi^2 M_n)$ in the bracket of first formula in (11) insignificantly changed results of calculations when holding the same value $G_e$. It seemingly happens due to the strong effect of stress induced crystallization in NR used in our tests.

**Conclusions**

The paper is based on the following concept. In soft materials like gels and cross-linked unfilled rubbers, the damage happens because of polymeric chains scission which changes molecular characteristics of damaged rubbers. This concept is quite different from the common view of damage in hard materials as related to occurrence of micro-cracks with unchanged material characteristics in undamaged regions (skeleton). This concept is also quite different from several models employed for describing the Mullins effect for rubbers. Unlike these models, the present model contains no fitting parameters for tensile tests, i.e. unknown parameters found while modeling the test data the model has to describe.

Using this concept, the damage model for cross-linked elastomers has been developed for slow loading when the rubber relaxations can be ignored. The general approach is based on the energy balance for chain scission (2), with an additional quasi-elastic description of damaged material. This general approach is further specified with (i) using the Gent elastic potential (4a) for undamaged materials and molecular evaluations of parameters in this potential proposed in this paper; (ii) rough assumptions of scaling evaluations (9) for change in molecular parameters in damaged rubber, and (iii) the formulation of terminal damage criterion (10) as a mechanical de-vulcanization in molecular terms. This criterion treats the failure in rubbers as a liquidation of cross-links, i.e. as a transition from cross-linked solid-like rubber to a highly branched liquid-like (green) rubber. Some experiments [43] with periodic impacts of a steel intruder imposed on a thick cross-linked rubber layer clearly demonstrated this type of damage as a formation of liquid rubber after many impacts. Since a speculative derivation of formula (10), some direct evaluations of cross-linked density using NMR, absent in our studies, will be highly appreciated.

Additionally, testing theory in tensile experiments for crystallizable rubbers like NR needed to involve in the model the effect of strain induced crystallization (SIC). Specific needle-like crystals formed in high stretching serve as additional cross-links delaying the failure [4]. Since the SIC theory was not well statistically developed we used in this paper the approximate approach of paper [32].

It should be noted that all currently popular formulations [29-31] of elastic CE's are based on entropic concept and ignore the enthalpic contribution in rubber elasticity. Therefore, these models cannot properly describe/predict either the stresses near the ultimate high strains and/or the typical enthalpic effects of scission of polymer chains related to distortion and dissociation of covalent bonds. That is seemingly the reason for deviation of model prediction and experimental data in Figure 4 for $\lambda > 6.5$. It seems that the more complicated hybrid theory [32] could be applied for resolving these problems, but damage model based on this theory has yet to be developed.

The approximate way of damage modeling in the paper, which involves a significant pre-damage sample deformation and determining the value $U_s$ in equations (1) or (2), is in accord with well known experimental fact that before scission, polymer chains should be highly extended. This fact was also recently confirmed in direct numerical computations [33].

The comparison of calculations with experiments shown in Figures 4-8 is encouraging. It seems that the damage of material semi-quantitatively predicted by our model could fairly describe the basic feature of mechanical behavior of NR rubber with no fitting parameters. Especially interesting are calculations of damage factor in consecutive extensions shown in Fig.8, which demonstrates that the initial high decrease in value of $\chi$ at the second cycle of extension changes to saturation when $\chi$ value approaches to its limit value at break. These calculations also present indirect confirmation of our scaling relations (9).

The problem of formation and accumulation of microscopic voids caused by the scission of polymeric chains has not been resolved in this paper. Our attempts to model the relation between the scission of polymer chains and occurring micro-voids failed. Experimental studies of testing the Zhurkov theory of life prediction for highly oriented polymers showed that this dependence is not easy to discover even in experiments (see the relevant discussions of contradictory viewpoints on the topic in the text [6]). Our observation and rough measurements show that the macroscopic cracks caused by growing these voids occur just before the rupture of the samples and insignificantly (less than few percents) increase the rubber specific volume. The comparison of our calculations with experiments showed that the real breakage of sample due to cracking happens earlier than the complete mechanical de-vulcanization of elastomers predicted by formula (10). Yet, this mechanical failure occurs in inhomogeneous regions of strains and stresses very close to the holders, seemingly due to the stress concentration effect, while our calculations assumed the homogeneous stress and strain fields in the tensile tests.

Finally, the relaxation properties of the cross-linked rubbers might also be easily incorporated in the approach elaborated in this paper as proposed in Refs.[42,43], if the dependence of linear relaxation spectrum on molecular parameters is known. However, in spite of general success in understanding the molecular mechanisms of cross-linked rubber relaxations [41], the linear relaxation spectrum has not yet been described in the molecular terms.

**Figure Captions**

Figure 1: Sketches of molecular modeling of rubbers: a) coil-like strands in unloaded rubber, b) alternative "blob" approach.

Figure 2: Figure 2: O-ring extended by steel hooks in experimental setup for tensile tests.

Figure 3: Repeating quasi-static loadings of NR O-ring samples for different extensional ratios: a) three repeats for final extension with $\lambda = 3.5$; b) three repeats for final extension with $\lambda = 4.5$.

Figure 4: Plot $\sigma(\lambda)$ in the quasi-static extension of fresh material till break; solid line – experimental data, line marked by symbols – calculations.

Figure 5: Calculated damage factor $\chi(\lambda)$ in the quasi-static extension shown in Fig.4

Figure 6: First quasi-static extension of fresh sample till $\lambda_1 \approx 6$; solid line –experimental data, line marked by symbols – calculations.

Figure 7: The second cycle of quasi-static extension till $\lambda_2 \approx 5.8$; solid line –experimental data, line marked by symbols – calculations.

Figure 8: Calculated plots $\chi(\lambda)$ in seven consecutive extension cycles up to $\lambda \approx 6$

**Figures**

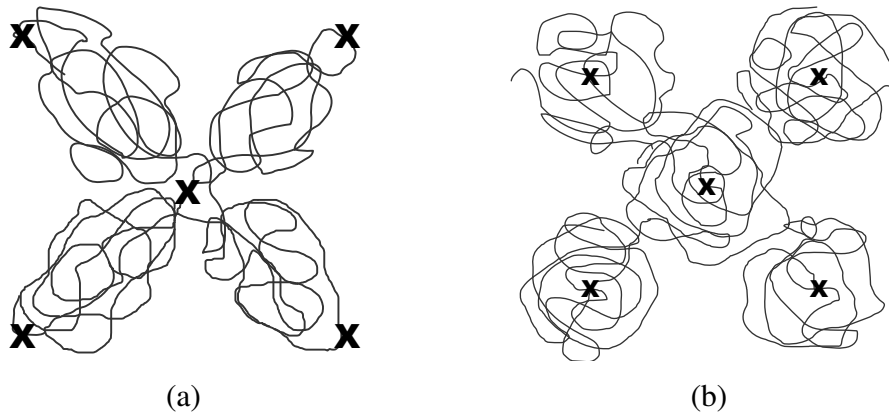

(a)             (b)

Figure 1: Sketches of molecular modeling of rubbers: a) coil-like strands in unloaded rubber, b) alternative "blob" approach.

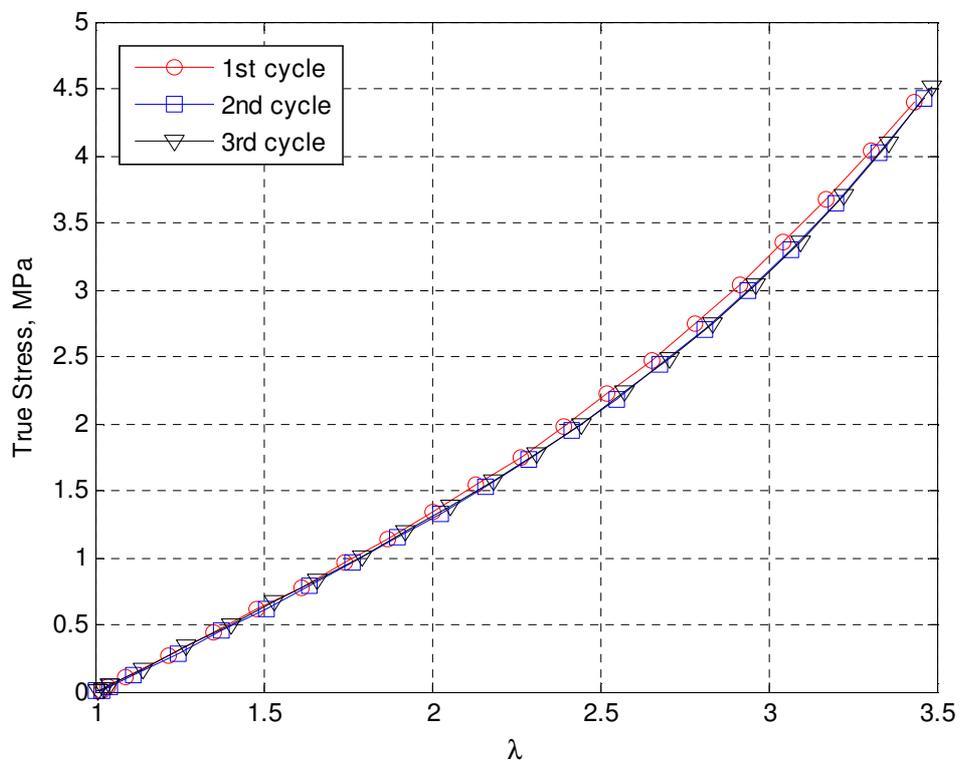

Figure 2a: Three quasi-static repeats for final extension with $\lambda = 3.5$.

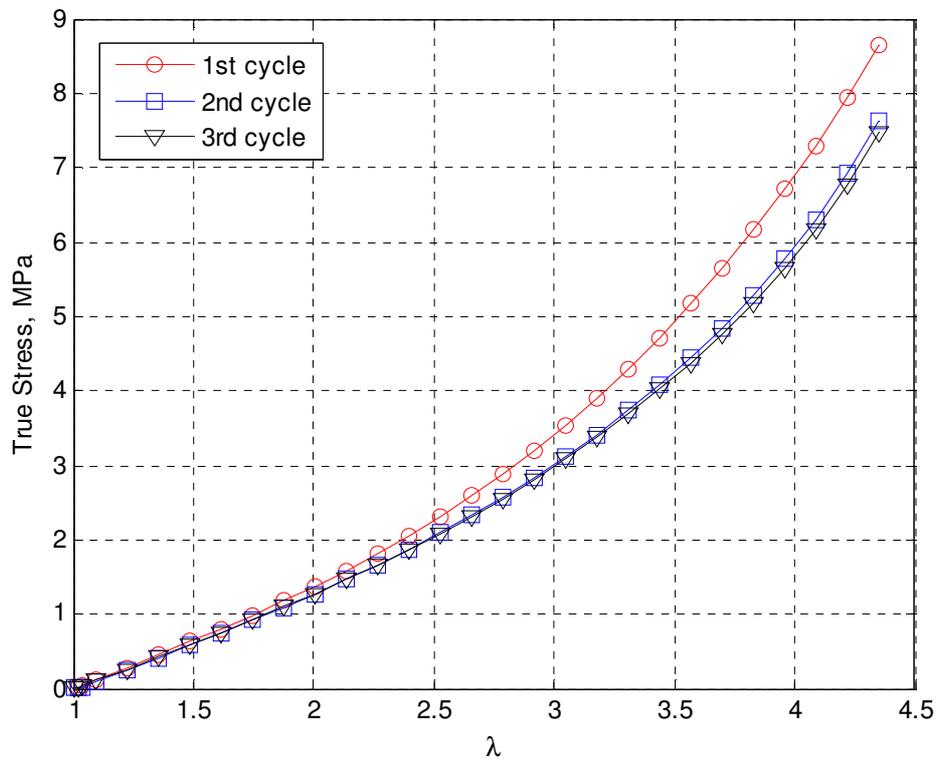

Figure 2b: Three quasi-static repeats for final extension with $\lambda = 4.5$.

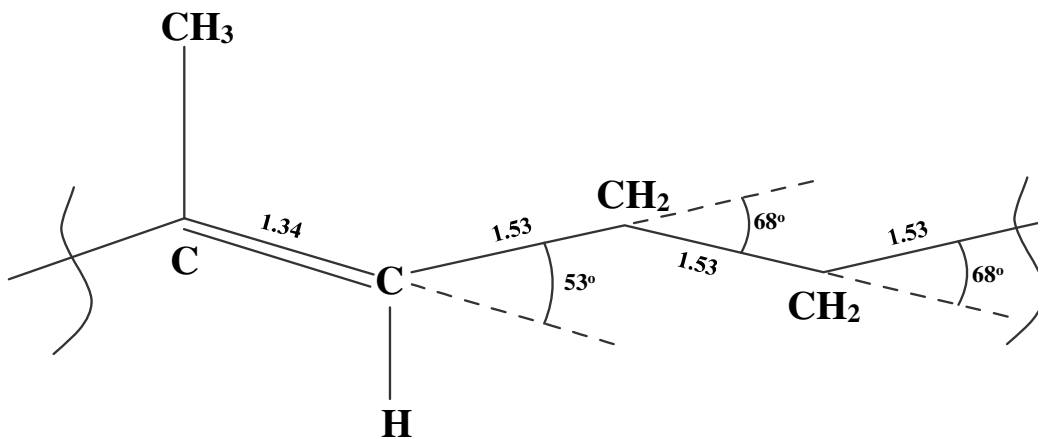

Figure 3: Structure and dimensions of monomer unit of *cis*-1,4-polyisoprene [42].

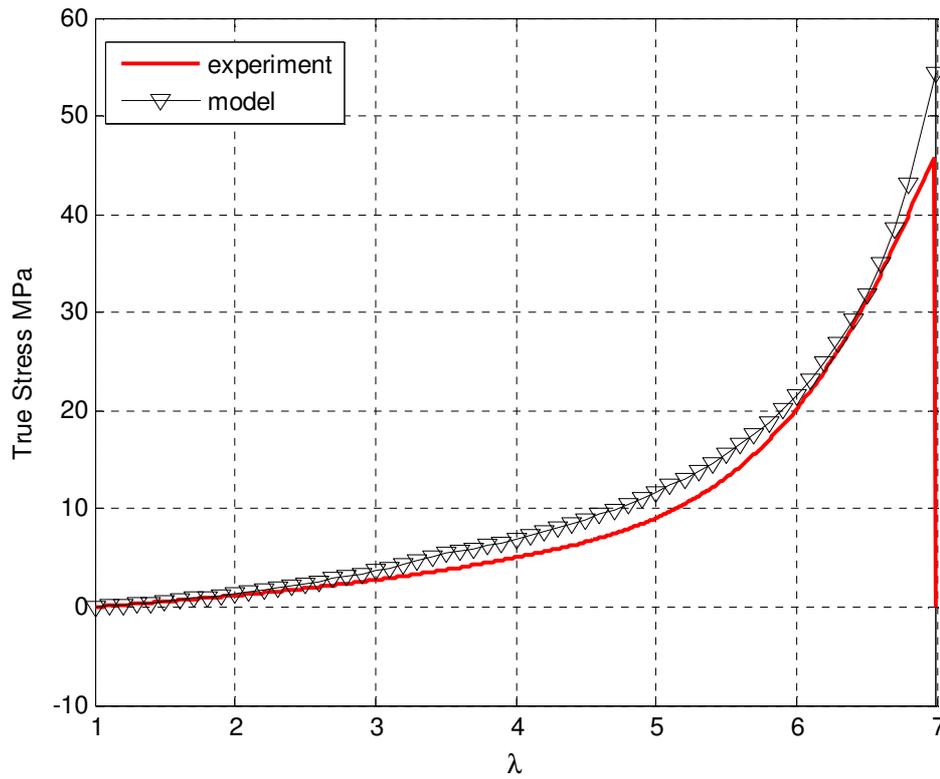

Figure 4: Plot $\sigma(\lambda)$ in the quasi-static extension of fresh material till break.

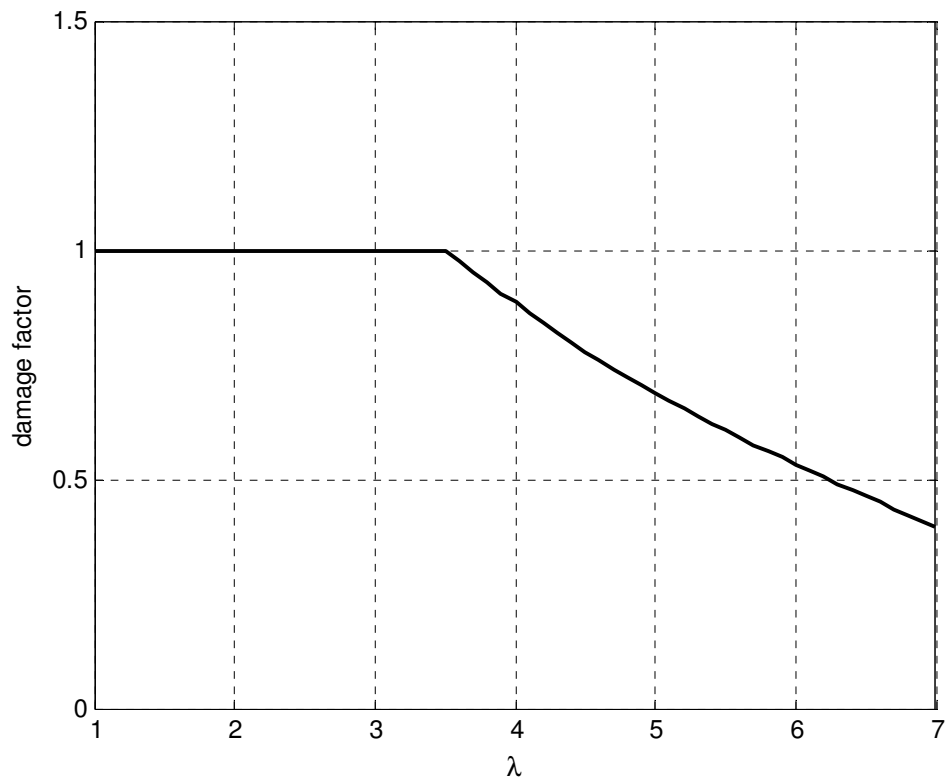

Figure 5: Calculated damage factor $\chi(\lambda)$ in the quasi-static extension shown in Fig.4

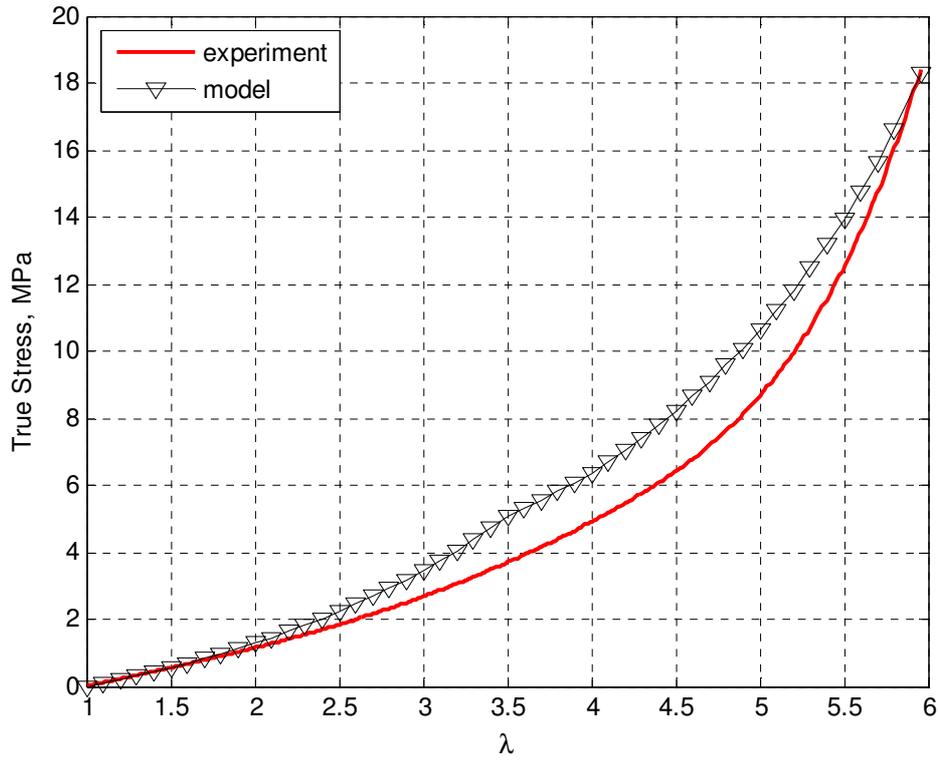

Figure 6: First quasi-static extension of fresh sample till $\lambda_1 = 6$.

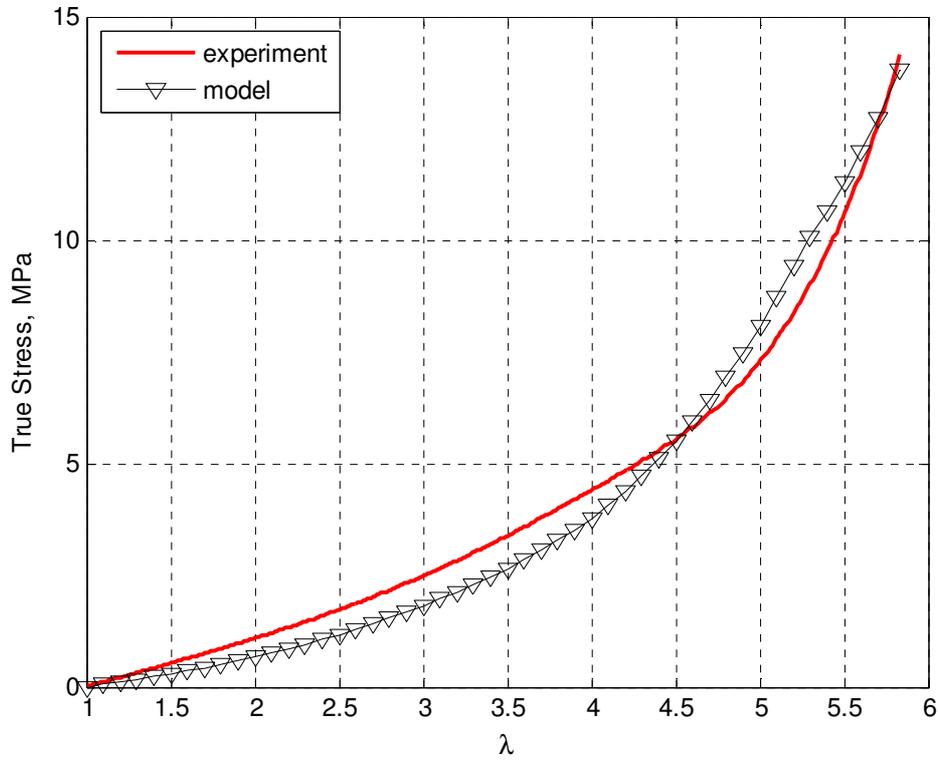

Figure 7: The second cycle of quasi-static extension till $\lambda_2 = 5.8$.

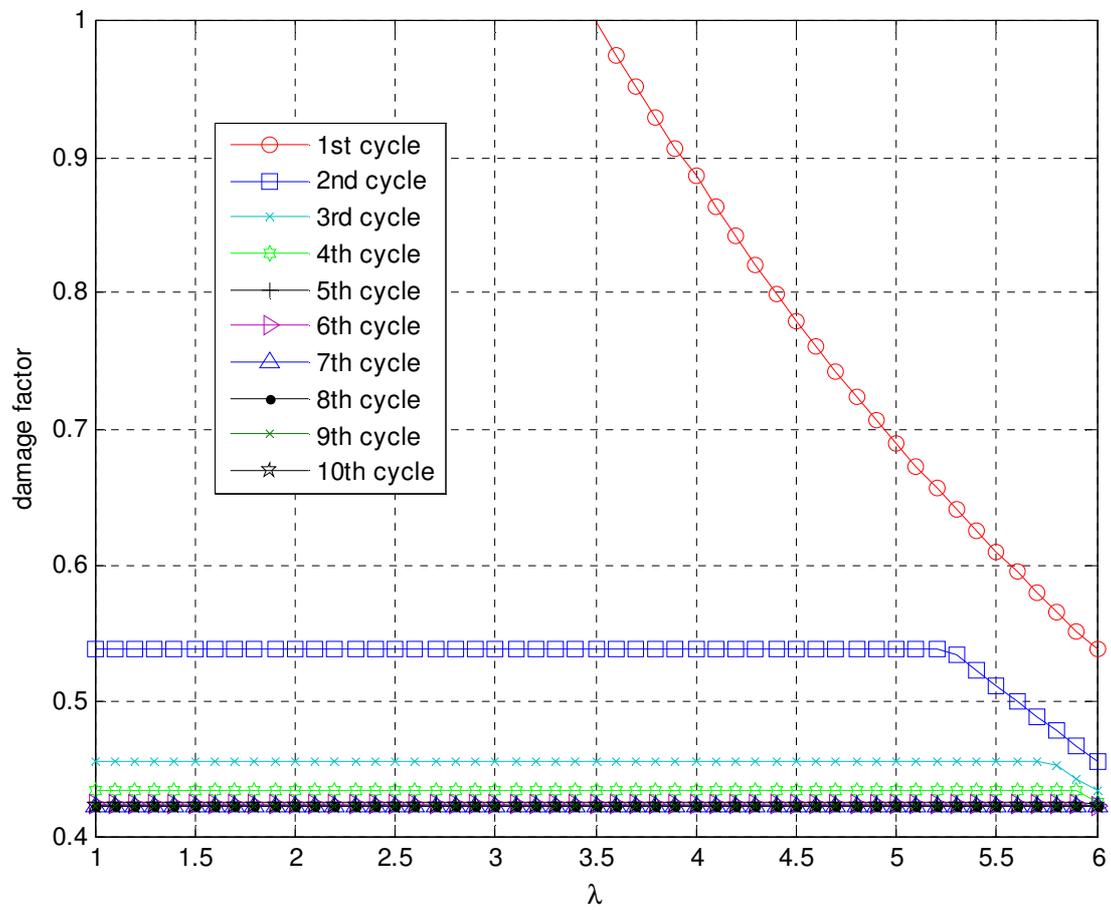

Figure 8: Calculated plots $\chi(\lambda)$ in seven consecutive extension cycles up to $\lambda \approx 6$